\documentstyle[aps,prb,amssymb,multicol,floats,epsf]{revtex}
\draft
\newcommand{\frontmatter}[1]{%
\twocolumn[
\hsize\textwidth\columnwidth\hsize\csname @twocolumnfalse\endcsname{#1}
\vspace*{0.5pc}]}

\title{Coulomb gap in a model with finite charge transfer energy.
}

\author{S.\ A.\ Basylko$^1$, P.\ J.\ Kundrotas$^{2,3}$,
V.\ A.\ Onischouk$^{1,2}$, E.\ E.\ Tornau$^{2,4}$ and 
A.\ Rosengren$^2$
}

\address{$^1$Joint Institute of Chemical Physics of Russian Academy 
of Sciences, 117 977 Kosygin Str.\ 4, Moscow, Russia
}
\address{$^2$ Department of Physics/Theoretical physics, Royal Institute
of Technology, SE--100 44 Stockholm, Sweden
}
\address{$^3$ Faculty of Physics, Vilnius University, Sauletekio al.\ 9,
LT--2040, Vilnius, Lithuania 
}
\address{$^4$Semiconductor Physics Institute, Go\v{s}tauto 11, 
LT--2600 Vilnius, Lithuania  
}

\date{Received\hspace*{5cm}}

\begin{document}
\frontmatter{
\maketitle

\begin{abstract}
The Coulomb gap in a donor-acceptor model with finite charge transfer
energy $\Delta$ describing the electronic system on the dielectric
side of the metal-insulator transition is investigated by means of
computer simulations on two- and three-dimensional finite samples with
a random distribution of equal amounts of donor and acceptor
sites. Rigorous relations reflecting the symmetry of the model
presented with respect to the exchange of donors and acceptors are
derived. In the immediate neighborhood of the Fermi energy $\mu$ the
the density of one-electron excitations $g(\varepsilon)$
is determined solely by finite size effects and $g(\varepsilon)$
further away from $\mu$ is described by an asymmetric power law with a
non-universal exponent, depending on the parameter $\Delta$.

\end{abstract}

\bigskip


}
\section{INTRODUCTION}

Doping of solids might lead to drastic qualitative changes in their
properties. The metal-insulator transition (MIT) is a spectacular
manifestation of this.  The understanding of the driving forces of the
MIT is a long-standing problem. In the early seventies, the prediction
\cite{pollak70} was made that on the dielectric side of the MIT the
long-range Coulomb interactions deplete the density of one-electron
excitations (DOE) $g(\varepsilon)$ near the Fermi energy
$\mu$. Further, analytical calculations of $g(\varepsilon)$ with
Coulomb correlation taken into consideration have been performed on
the metallic side of the MIT. Altshuler and Aronov \cite{alts79}
showed that for the metallic case $g(\varepsilon)$ in three
dimensions has a cusp-like dependence $g(\varepsilon)\sim |\varepsilon
- \mu|^{1/2}$ near $\mu$. This was later confirmed in electron
tunneling experiments for amorphous alloys \cite{mcmil81} and granular
metals \cite{imry85}.

On the insulating side of the MIT charge transport occurs via
inelastic electron tunneling hopping between states localized on the
impurity sites with one-electron energies close to $\mu$. Mott
\cite{mott68} demonstrated that at low temperatures electrons seek
accessible energy states by hopping distances beyond the localization
length, leading to a hopping conductivity $\sigma(T)\sim \exp
(-T_0/T)^\nu$ with $T_0$ being a characteristic temperature depending
on localization length and with the hopping exponent $\nu=1/4$ for the
non-interacting case in three dimensions. Efros and Shklovskii
\cite{efros75} (ES) argued that the ground state of a system with
long-range Coulomb interactions is stable with respect to one-particle
excitations only if $g(\varepsilon)$ in the vicinity of $\mu$ has the
symmetric shape
\begin{equation}
g(\varepsilon)\sim |\varepsilon - \mu|^{D-1}
\label{efros}
\end{equation} 
with the universal exponent $D-1$ depending only on the dimensionality
$D$ of the system. In particular, ES predicted that in $D=3$
$g(\varepsilon)=\frac{3}{\pi}\left( \frac{\chi}{e^2}\right) ^3
(\varepsilon - \mu)^2$,
where $\chi$ is the dielectric constant and $e$ is the electron
charge. Because $g(\varepsilon)$ vanishes only at $\varepsilon=\mu$,
this is called a ``soft'' Coulomb correlation gap with a width
$\Delta \varepsilon \sim e^3(N_0/\chi ^3)^{1/2}$, where $N_0$ is
the DOE far away from $\mu$. The power law (\ref{efros}) gives
\cite{pollak72} a hopping exponent $\nu=D/(D+3)$ at low
temperatures, so for three-dimensional system with long-range Coulomb
interactions $\nu=1/2$.

The intriguing hypothesis about universality of (\ref{efros}) has
stimulated further theoretical research, both
analytical \cite{bar80} and numerical
\cite{bar79,efros79,dar84,mob88,ten93}. 
To establish the hypothesis (\ref{efros}) Efros\cite{efros76} used 
the ground-state stability conditions for localized electrons (LES)
with respect to charge transfer
\begin{equation}
\varepsilon_j - \varepsilon_i -\frac{e^2}{\chi r_{ij}} > 0,
\label{stabcon}
\end{equation}
where $\varepsilon_i$ and $\varepsilon_j$ are the one-particle
energies of a neutral donor on a site $i$ and of a charged donor on a
site $j$, respectively, and $r_{ij}$ is the distance between the sites
$i$ and $j$.  The conditions (\ref{stabcon}) were used to
heuristically derive a non-linear integral equation for
$g(\varepsilon)$ \cite{efros76,efros84,burin95,mog89} and then
assymptotic analysis of this equation leads \cite{efros76} to the
power law (\ref{efros}).

LES have been studied using the so-called classical donor-acceptor
($d$-$a$) model (see, e.\ g.\ Ref.\onlinecite{efros84}). Within this
model, the system considered is modeled by a continuous sample with
randomly distributed $k\times N$ ($k\leq 1$) acceptor and $N$ donor
sites. Each acceptor site is negatively charged whereas out of $N$
only $k\times N$ donors have a positive charge which leads to a large
number of configurations of charged donors. Moreover, each of these
configurations must obey not only conditions (\ref{stabcon}) but also
more complicated conditions related to many-particles excitations (e.\
g., charge transfer involving four, six, etc. sites). Efros conjecture
about the {\it universality} implies that $g(\varepsilon)$ does not
depend on peculiarities of the particular model and, as a consequence,
further theoretical studies of LES \cite{bar79,dar84,mob88,ten93} were
confined to a {\it lattice} $d$-$a$ model proposed in
Ref.\onlinecite{efros76}.  In this model, $N$ donors are localized on
all the sites of a D-dimensional lattice and the negative charge from
$k\times N$ acceptors is uniformly smeared over the lattice sites so
that each site $i$ has a charge $e(n_i-k)$, where $n_i=1$ if a donor
on the site $i$ is ionized and $n_i=0$ if a donor is neutral. Disorder
in this model is ensured by introducing randomly distributed one-site
potentials. Monte Carlo simulations \cite{mob88} on very large
specimens of the lattice $d$-$a$ model, however, have given rise to
doubts about the universality of the $g(\varepsilon)$ behavior.

Another hint about possible non-universal behavior of $g(\varepsilon)$
has come from the intriguing and still not completely unfolded problem
whether the so called spin-glass phase does exist in the classical
$d$-$a$ model (see, e.\ g.\
Ref.\onlinecite{gran93,lik98,voj94}). Grannan and Yu \cite{gran93}
studied the classical three-dimensional $d$-$a$ model with $k=0.5$ but
with the total acceptor charge uniformly distributed over donor sites
as in the lattice $d$-$a$ model. In this case, the classical $d$-$a$
model is equivalent to a model of Ising spins, localized on randomly
distributed sites, with pairwise Coulomb interactions, a model in
which a transition into the spin-glass state was found \cite{gran93}
to occur at non-zero temperature. It was then concluded that such a
transition should exist in {\it all} $d$-$a$ models (with and without
smearing of negative charge, defined on a lattice or on a continuous
sample) as well because of the Efros universality hypothesis. Voita and Schreber
\cite{voj94}, however, have shown that the spin glass transition does
not exist in the lattice $d-a$ model\cite{efros76}. Besides, in recent
work by one of us \cite{lik98} it was unequivocally demonstrated that
the ground state of the classical $d$-$a$ model and that of the model
studied in Ref.\onlinecite{gran93} are qualitatively different. An
analysis of histograms ${\cal H}[Q_{\alpha \beta}]$ of the so called
overlaps $Q_{\alpha \beta}=\frac{1}{N}\sum_i
\delta(n_i^{\alpha},n_i^{\beta})$ (here $\alpha$ and $\beta$ refer to
different pseudo-ground states (PGS) obtained by direct descents) has
revealed that, indeed, for the model studied in
Ref.\onlinecite{gran93} ${\cal H}[Q_{\alpha \beta}]$ has a symmetric
Gaussian shape with the maximum at $\langle Q_{\alpha \beta}\rangle
=0$ and with the dispersion $\langle Q_{\alpha \beta}^2\rangle \sim
N^{-1}$. This means that a large number of microscopically different
PGS's does exist in the model and according to Parisi's theory
\cite{mez87} this implies the existence of a spin-glass state at low
temperatures. Further Monte Carlo simulations at finite temperatures
\cite{lik98} revealed the typical finite-size scaling of the
spin-glass susceptibility. In the classical $d$-$a$ model, however,
${\cal H}[Q_{\alpha \beta}]$ has its maxima at $\langle Q_{\alpha
\beta}\rangle =1$ which means that all PGS's generated are the same
from microscopical point of view.  The absence of microscopically
different PGS's in the classical $d$-$a$ model was explained
\cite{lik98} by the pinning of all PGS's on the electric field created
by the discretely distributed acceptor charges.

Therefore, it is highly desirable to study the properties of not only the
classical $d$-$a$ model, but of its various modifications as well. In
the present work we consider a modified classical $d$-$a$ model
(MCDAM) in which acceptors can be neutral, so the energy $\Delta$ of
the charge transfer from a donor to an acceptor ($d^0+a^0\rightarrow
d^+ + a^-$, where $d^0$, $d^+$, $a^0$, $a^-$ stand for a neutral
donor, a charged donor, a neutral acceptor, a charged acceptor,
respectively) has to be finite. The classical $d$-$a$ model might be
then viewed as the limit of the MCDAM as $\Delta\rightarrow \infty$. We have
investigated the shape of the Coulomb gap (i.\ e.\ $g(\varepsilon)$
both for donor and acceptors in the vicinity of the Fermi level) in
two- and three-dimensional MCDAMs at T=0 and found that the behavior of
$g(\varepsilon)$ is in strong contradiction to the Efros conjecture
about the universality of $g(\varepsilon)$. The rest of the paper is
organized as follows. In Section II we introduce the MCDAM and arrive
at some rigorous results which follow from a symmetry of the MCDAM
with respect to the exchange of donor and acceptor sites. Further,
the algorithm of energy minimization for the MCDAM including a discussion
about inherent finite size effects is presented in Section
III. Section IV is devoted to a description of the main results
obtained. In Section V we discuss possible causes of universality
violation in the MCDAM, analyze experimental data available in the
literature and predict possible experimental situations in which
the non-universal behavior of $g(\varepsilon)$ might be observed. And
finally, a summary is presented in Section VI.

\section{BACKGROUND}

\subsection{Model}

We consider a $D$-dimensional system of volume $L^D$, in which an equal
number of acceptor and donor sites $N$ are allocated according to the
Poisson distribution with a density $n=N\times L^{-D}$. It is
convinient to choose energy unit $E_0$ as an energy of the Coulomb
interaction between a pair of acceptors, say, localized on the average
distance $n^{-1/D}$, $E_0=e^2 n^{1/D}/\chi$. In typical bulk
semicondustors $n \sim 10^{18}$ cm$^{-3}$ and $\chi \sim 10$, so
$E_0\sim 0.02$ eV. Hereafter all expressions will be written in
dimensionless units $n^{-1/D}$ for length and $E_0$ for energies.  A
microscopic state of a particular spatial arrangement of the donor and
acceptor sites (henceforth referred to as the sample {\bf R}) is
determined by a set of occupation numbers $(n_a,n_d)\equiv
\{n_a(i),n_d(k), i=1,2,\ldots ,N$, $k=1,2,\ldots ,N\}$ determined in
the following way. For the acceptors, $n_a(i)=1$ if an acceptor on an
acceptor site $i$ has captured an electron and $n_a(i)=0$ if an
acceptor is neutral. For the donors, $n_d(k)=1$ if a donor on a donor
site $k$ is neutral and $n_d(k)=0$ if a donor has given an electron
away. We investigate the LES from the dielectric side of the MIT, so
$\alpha_B<1$ ($\alpha_B$ is the localization length of the electron on
donor).  The energy of the sample, assuming that all the interactions
are of Coulomb origin, then is
\begin{eqnarray}
E(n_a,n_d)& = &\frac{1}{2}\; \sum_{i\ne j} \frac{n_a(i)\; n_a(j)}
{r_{ij}^{a-a}} + \nonumber \\
&+&\frac{1}{2} \; \sum_{k\ne l} \frac{(1-n_d(k))\; (1-n_d(l))}
{r_{kl}^{d-d}}- \nonumber \\
&-&\; \sum_{i,k} \frac{(1-n_d(k))\;n_a(i)}
{r_{ik}^{a-d}}- \Delta \sum_i n_a(i),
\label{energy}
\end{eqnarray}
where indices $i,j$ and $k,l$ number acceptor and donor sites,
respectively, $r_{ij}^{a-a}$, $r_{kl}^{d-d}$ and $r_{ik}^{a-d}$ are
the distances between the acceptors on the sites $i$ and $j$, between
the donors on the sites $k$ and $l$, and between the acceptor on the
site $i$ and the donor on the site $k$, correspondingly, and $\Delta$
is the energy of charge transfer between acceptor and donors. When
charge transfer occurs in the system, the energy of the sample changes
by
\begin{eqnarray}
\delta E(n_a,n_d)& = &\sum_{i} \varepsilon_a(i)\delta n_a(i)+
\sum_{k} \varepsilon_d(k)\delta n_d(k)+\nonumber \\
&+&\sum_{i,k} \frac{\delta n_a(i)\; \delta n_d(k)}{r_{ik}^{a-d}}
+\frac{1}{2}\; \sum_{i\ne j} \frac{\delta n_a(i)\; 
\delta n_a(j)}{r_{ij}^{a-a}}+\nonumber \\
&+&\frac{1}{2}\; \sum_{k\ne l} \frac{\delta n_d(k)\; 
\delta n_d(l)}{r_{kl}^{d-d}},
\label{change}
\end{eqnarray}
where $\varepsilon_a(i)$ is the one-electron excitation (OEE) energy
for the acceptors
\begin{equation}
\varepsilon_a(i)\; \equiv \; \frac{\delta E(n_a,n_d)}{\delta n_a(i)}\; =\; 
\sum_{j\ne i} \frac{n_a(j)}{r_{ij}^{a-a}}-\sum_k \frac{1-n_d(k)}
{r_{ik}^{a-d}} - \Delta,
\label{eps}
\end{equation}
 $\varepsilon_d(i)$ is the corresponding OEE energy for the donors and
$\delta n_a(i)$ ($\delta n_d(k)$) denotes the change of the occupation
number on the acceptor (donor) site $i$ ($k$). If a microscopic state
($n_a^0$,$n_d^0$) of the sample is the ground-state of this sample
then for any excitation the relation
\begin{equation}
\delta E(n_a^0,n_d^0)\geq 0
\label{ground_state}
\end{equation}
holds. The specific appearance of the conditions (\ref{ground_state})
depends on what excitations are allowed in the model system considered.

In the present paper we investigate the simplest case when only pairs
of sites are involved in the charge transfer which, in turn, is
allowed to occur in four different ways: (i) via electron hops
between a pair of the acceptors $\{n_a(i)=1,n_a(j)=0\}\rightarrow
\{n_a(i)=0,n_a(j)=1\}$; (ii) via electron hops between a pair of
donors $\{n_d(k)=1,n_d(l)=0\}\rightarrow \{n_d(k)=0,n_d(l)=1\}$; (iii)
via ionization process $\{n_a(i)=0,n_d(k)=1\}\rightarrow
\{n_a(i)=1,n_d(k)=0\}$ and (iv) via recombination process
$\{n_a(i)=1,n_d(k)=0\}\rightarrow \{n_a(i)=0,n_d(k)=1\}$. For each of
those processes there is an unique set of $\{ \delta n_a(i),\delta
n_a(j),\delta n_d(k),\delta n_d(l)\} $. For instance, for the
acceptor-acceptor hops
\begin{equation}
\delta n_a(i) = -1,\; \delta n_a(j)=1,\;
\delta n_d(k)=0,\; \delta n_d(l)=0.
\label{daa}
\end{equation}
Substituting (\ref{daa}) into (\ref{change}) one obtains the ground-state
stability relation with respect to the charge transfer between the
pair of acceptors on the sites $i$ and $j$
\begin{equation}
\varepsilon_a^0(j) -\varepsilon_a^1(i)-\frac{1}{r_{ij}^{a-a}}\geq 0,
\label{aa}
\end{equation}
where $\varepsilon_a^{1(0)}(i)$ denotes $\varepsilon_a(i)$ if
$n(i)=1(0)$.  The stability conditions with respect to the other three
manners of the charge transfer are obtainable in the similar manner.

The relation (\ref{aa}) implies that $\varepsilon_a$'s for the neutral
acceptors are, in general, larger than $\varepsilon_a$'s for the
charged acceptors. Furthermore, the pair of neutral and charged
acceptors might be located on any distance and therefore in the
thermodynamic limit the chemical potential for the acceptors (i.\
e.\ an energy level which separates the energies of the neutral
and charged acceptors) is determined as
\begin{equation}
\mu_a=\text{min}\{\varepsilon_a^0(i)\}
=\text{max}\{\varepsilon_a^1(i)\}.
\label{chempot}
\end{equation}
Alike, there exist the chemical potential $\mu_d$ for the donors as
well. Moreover, the stability relations with respect to the ionization
and recombination lead to 
\begin{equation}
\mu_a=\mu_d=\mu.
\label{muequal}
\end{equation}
Despite the finite size of samples we investigated, the relation
(\ref{muequal}) with the chemical potentials calculated from
(\ref{chempot}) is valid within the limits of accuracy of our
calculations (see Sect. III).

A macroscopic state of the sample {\bf R} is characterized by degree of
acceptor ionization
\begin{equation}
C_a({\bf R})=\frac{1}{N}\sum_i n_a(i),
\label{ioniz}
\end{equation}
by the DOE for acceptors
\begin{equation}
g_{a}(\varepsilon_a ,{\bf R})=\frac{1}{N}\sum_i \delta(\varepsilon -
\varepsilon_a(i))
\label{doe}
\end{equation}
and by the corresponding DOE $g_{d}(\varepsilon_d ,{\bf R})$ for the
donors.  Note, that for the finite samples (especially for the
relative small systems we were able to investigate) $C_a({\bf R})$,
$g_{a}(\varepsilon_a ,{\bf R})$ and $g_{d}(\varepsilon_d ,{\bf R})$
depend essentially on the particular implementation {\bf R} of the spatial
distributions of the donor and acceptor sites (if a sample would be big
enough all quantities would be self-averaging). Therefore, in order to
obtain reliable results, one has to work with the quantities
$C_a\equiv \langle C_a({\bf R})\rangle $, $g_a(\varepsilon)\equiv
\langle g_{a}(\varepsilon_a ,{\bf R})\rangle $ and
$g_d(\varepsilon)\equiv \langle g_{d}(\varepsilon_d ,{\bf R})\rangle
$, where $\langle \ldots \rangle$ denotes the average over a number of
{\bf R}'s.  Note, that the values $g_{a(d)}(\varepsilon_{a(d)},{\bf
R})d\varepsilon$ obtained for independent {\bf R}'s are scattered
according to the Gaussian distribution with the mean
$g_{a(d)}(\varepsilon)d\varepsilon$ and the standard deviation
$\sqrt{g_{a(d)}(\varepsilon)d\varepsilon}$. In the region of the
Coulomb gap $g_{a(d)}(\varepsilon)d\varepsilon \sim 10^{-4}$ and
dispersion is several orders of magnitude larger than the
mean. Therefore, in order to reduce the statistical noise in the final
$g_{a(d)}(\varepsilon)$ dependences an average is needed over
a sufficient large amount of independent samples (we performed
calculations with up to 10$^4$ samples).

\subsection{Acceptor-donor symmetry}

Let us rewrite the energy (\ref{energy}) in terms of the 
OEE energies (\ref{eps})
\begin{eqnarray}
E(n_a,n_d)&=&\frac{1}{2}\; \sum_i \varepsilon_a(i)n_a(i)-\nonumber \\
&-&\frac{1}{2}\; \sum_k \varepsilon_d(k)(1-n_d(k))- 
\frac{\Delta }{2}\; \sum_i n_a(i).
\label{energy1}
\end{eqnarray}
The system investigated is electrically neutral, i.\ e. for any sample
\begin{equation}
\sum_i n_a(i)= \sum_k (1-n_d(k)).
\label{neutral}
\end{equation}
Then, the energies of the microscopic states ($n_a, n_d$) and
($n_a^*, n_d^*$) for a sample {\bf R} and its ``mirror''
reflection {\bf R$^*$} (when the donor and acceptor sites exchange
places keeping the spatial arrangement of sites unchanged), are equal
under the following conditions
\begin{equation}
\varepsilon_a(i)+\varepsilon_d^*(i)=
\varepsilon_d(k)+\varepsilon_a^*(k)=-\Delta
\label{symm1}
\end{equation}
and
\begin{equation}
n_a^*(i)=(1-n_d(i))\hspace{1cm}n_d^*(k)=(1-n_a(k)).
\label{symm2}
\end{equation}
The stability relations (\ref{aa}) for the ground-state
($n_a^0$,$n_d^0$) of the sample {\bf R} transform into stability
relations for the ground-state ($n_a^{0*}$,$n_d^{0*}$) of the sample
{\bf R$^*$} through the relations (\ref{symm1},\ref{symm2}) as well.

Since averaging over samples includes all possible pairs {\bf R} and
{\bf R$^*$}, it follows from the symmetry relations (\ref{symm1}) and
(\ref{symm2}) along with the definition (\ref{doe}) that
$g_a(\varepsilon)$ can be mapped to $g_d(\varepsilon)$ using the
relation
\begin{equation}
g_d(\varepsilon)=g_a(-\varepsilon-\Delta)
\label{mapp}
\end{equation}

The symmetry of the model imposes also a relation between the Fermi
energy $\mu$ (\ref{chempot},\ref{muequal}) and the parameter
$\Delta$ of the model. Expressing $n_{a[d]}(i[k])$ in terms of the
Heaviside's step functions $n_{a[d]}=\theta(\mu -
\varepsilon_{a[d]}(i[k]))$, the quantity $C_a$ can be written in the
form
\begin{equation}
C_a=\int_{-\infty}^{\mu} g_a(\varepsilon)d \varepsilon=
\int_{\mu}^{\infty} g_d(\varepsilon) d\varepsilon
\label{integ1}
\end{equation}
The symmetry relation (\ref{mapp}) transforms (\ref{integ1}) into an
integral relation
\begin{equation}
\int_{-\mu -\Delta}^{\infty} g_d(\varepsilon)d \varepsilon=
\int_{\mu}^{\infty} g_d(\varepsilon) d\varepsilon,
\label{integ2}
\end{equation}
which has a meaning only if
\begin{equation}
\mu=-\frac{\Delta}{2}\;. 
\label{mudelta}
\end{equation}
Thus, the Fermi energy of our model system in the
thermodynamic limit is a fundamental quantity depending only on the
energy of charge transfer from an acceptor to a donor.

\section{METHOD}

\subsection{Algorithm of energy minimization}

We start from a random allocation of $N$ donor and $N$ acceptor sites
in the continuous $D$-dimensional system (generate a sample {\bf R})
with the density $n=1$, so that the system has a linear size
$L=N^{1/D}$ and then charge randomly chosen $C_a\times N$ both donors
and acceptors (usually we take $C_a=0.7$), i.\ e.\ generate an initial
microscopic state (IMS) ($n_a$,$n_d$) of the sample {\bf
R}. Further, we search for such microscopic state ($n_a^0$,$n_d^0$)
which obeys the stability conditions (\ref{aa}) with respect to the
four mechanisms of the charge transfer allowed in our model. We used
an algorithm which is an extension of the algorithm proposed in
Ref.\onlinecite{bar79} to the case $\Delta \ne \infty$.  The algorithm
consists of the three main steps.

In order to save computer time, first, we look for pairs $a^0 - a^-$
($d^0 - d^+$) for which the ``crude'' stability relation $\Delta
\varepsilon \equiv \varepsilon_{a(d)}^0 - \varepsilon_{a(d)}^1>0$ is
violated. Then, the energy of the system is decreased by transferring
an electron between such pair of sites for which $\Delta \varepsilon$
has its minimal non-positive value. This process is repeated until a
state is reached, in which $\Delta \varepsilon > 0$ for all possible
$a^0 - a^-$ and $d^0 - d^+$ pairs (step I). In the similar manner, we
further minimize the energy of the system with respect to the ``true''
stability relations (\ref{aa}) for the charge transfer between the
$a^0 - a^-$ and $d^0 - d^+$ pairs (step II). And, finally, in the step
III we diminish the energy of the system with respect to the stability
relations for ionization and recombination processes. Since ionization
and recombination processes change the degree $C_a$ of the system
ionization, each time after one of these processes takes place during
calculations, we go back to the step II.  Repeating the steps II and
III, we finally arrive at a microscopic state
($n_a^0$,$n_d^0$) for which all four stability conditions are
fulfilled. We name the procedure ($n_a$,$n_d$)$\rightarrow$
($n_a^0$,$n_d^0$) via above steps I,II and III as ``a single descent''.

It should be noted, however, that the state ($n_a^0$,$n_d^0$) is not
necessarily the ground state of the sample {\bf R} since for the ground
state, in general, not only the simplest relations (\ref{aa}) with only
pairs of sites included, but the more complicated relations involving
quadruplets, sextets, etc. of sites have to be fulfilled. Therefore,
the state ($n_a^0$,$n_d^0$) (after Ref.  \onlinecite{bar79}) hereafter
will be referred to as the pseudo-ground state (PGS) of the sample
{\bf R}. Then, two questions naturally arise: How close the PGS and the
ground state of the given sample are and how this may influence the 
output of our calculations? In order to answer the first question, we
calculate and analyze the histograms $\cal{H}$ for the so-called overlaps
\begin{equation}
Q_{\alpha \beta} = \frac{1}{N}\sum_i \delta (n_a^{\alpha},n_a^{\beta}),
\label{qab} 
\end{equation}
where indices $\alpha$ and $\beta$ refer to PGS's which are obtained
by means of the single descent on the same sample but with different
IMS ($n_a$,$n_d$). If two PGS's are identical then $Q_{\alpha
\beta}=1$. We calculated for the $D=2$ system with $N=500$ at
$\Delta=0$ the mean $Q({\bf R})= \langle Q_{\alpha
\beta}\rangle_{\alpha \beta}$ for the sequence of 100 PGS's generated
by single descents from the different IMS of the same sample {\bf
R}. We further acquire $Q({\bf R})$ for 100 different samples and
obtain that the mean $\bar{Q}\equiv \langle Q({\bf R})\rangle_{\bf
R}=0.96$. It means that in PGS generated by the single descent only 20
acceptors out of 500 are, in average, in the ``wrong'' states compared
to those in the true ground state of the sample.

In order to evaluate how the ``erroneousness'' of PGS influences the
outcome of our calculations we perform an analysis of ground states
obtained by means of the so called multirank descents. Descent of rank
$m$ comprises of a consequence of the single descents on the same
sample with different IMS when calculations are stopped after the
lowest observed PGS energy repeats $m$ times. We calculate $\bar{Q}$
(all other parameters were the same as described in the previous
paragraph, where actually the case $m=0$ was explored) for descents
with different ranks $m=5,10,15$ and found that, for instance, for
$m=15$ (which implies drastic increase in the computation time)
$\bar{Q}=0.990$. $g_a(\varepsilon)$ and $g_d(\varepsilon)$ obtained
from the PGS's generated by means of the single descents and by means
of descents with $m=10$, say, do not differ within the limits of
statistical errors. So, we conclude, that reliable results can be
obtained by means of single descents already, thereby saving a lot of
computer time and resources. 

\subsection{Finite-size effects}

Due to constraints in computer resources, the largest samples, we were
able to deal with, comprise up to $N=2000$ donor and $N=2000$ acceptor
sites ($L\sim 45$ for $D=2$ and $L\sim 12$ for $D=3$). Such relative
small sizes of the samples investigated might influence the outcome of
calculations. Detailed analysis of finite size effects on the results
obtained will be presented in Section IV and here we want to make
two remarks about inherent finite size effects in the model system
considered.

First, as follows from (\ref{aa}), the energies $\varepsilon_a^0$ for
the neutral acceptors and $\varepsilon_a^1$ for the charged ones in 
finite samples at $T=0$ cannot be further away than $(L\times \sqrt{D})^{-1}$.
This implies that $g(\varepsilon_a)=0$ within the $\varepsilon_a$ interval 
\begin{equation}
|\varepsilon_a-\mu |< (2L\times \sqrt{D})^{-1}
\label{finite1}
\end{equation}
Of course, the same holds for donors as well. The relation
(\ref{finite1}) gives the estimation how close to $\mu$ data on the
energy spectrum are, in principle, obtainable from the calculations
on finite samples.

Secondly, as follows from (\ref{eps}) the energies $\varepsilon_a$ and
$\varepsilon_d$ for the finite samples are sensitive to the
location of the donor and acceptor sites. Therefore, the Fermi
energy $\mu$ for finite samples does differ, in general, from
sample to sample. A straightforward averaging of $g(\varepsilon)$ over
different samples might thus lead to a distortion of the
$g(\varepsilon)$ shape especially in the region where the Coulomb gap
is observed. In order to avoid this undesired effect, we used a trick
first proposed in Ref.\onlinecite{bar79}. During accumulation of the
results for $g(\varepsilon)$ we added together $g(\varepsilon)$ for
the same values of $\varepsilon -\mu({\bf R})$ rather than for the
same values of $\varepsilon$. Here $\mu({\bf R})$ denotes the Fermi
energy for a finite sample {\bf R} calculated as
\begin{equation}
\mu({\bf R})\; =\; \frac{1}{2}\; \bigg( \text{min}\{\varepsilon_a^0(i)\}
+\text{max}\{\varepsilon_a^1(i)\}\bigg),
\label{mufinite}
\end{equation}
Such way of doing $g(\varepsilon)$ average entirely excludes the
influence of the fluctuations of the Fermi energy in the finite
samples on the shape of the Coulomb gap.

\begin{figure}
\epsfxsize=3.33in
\centerline{\epsfbox{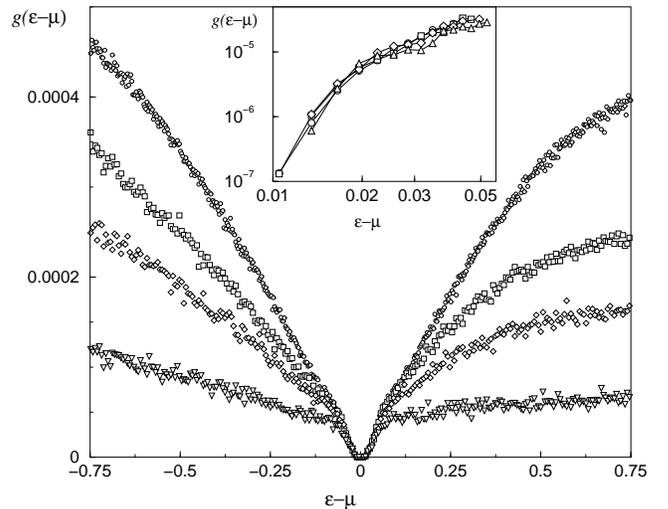}}
\caption{
Density of one electron excitations $g_a(\varepsilon-\mu)$ in
the vicinity of the Fermi energy $\mu$ obtained for
the two-dimensional model (\protect{\ref{energy}}) with $N=1500$ at
$\Delta=0$ (circles), $2$ (squares), $4$ (diamonds) and $10$
(triangles). Data points presented in the figure are calculated as
the average over 10.000 ($\Delta=0$), 5.100 ($\Delta=2$), 3.700
($\Delta=4$) and 2.200 ($\Delta=10$) different samples. Insert shows
double logarithmic plot of $g_a(\varepsilon-\mu)$ for $\varepsilon > \mu$
in the region $\varepsilon - \mu \lesssim 0.05$.
}
\label{fig1}
\end{figure}

Finally, we remark that all the data presented below were obtained
for the open boundary condition. In order to ensure that results
obtained are not determined by the type of the boundary conditions
used in calculations, we performed calculations of the two-dimensional
MCDAM at $\Delta=0$ with different $N$ and found that periodic boundary
conditions only effectively reduce the linear size of a sample, leaving
the qualitative shape of the parameters calculated unchanged.

\section{RESULTS}

\begin{figure}
\epsfxsize=3.33in
\centerline{\epsfbox{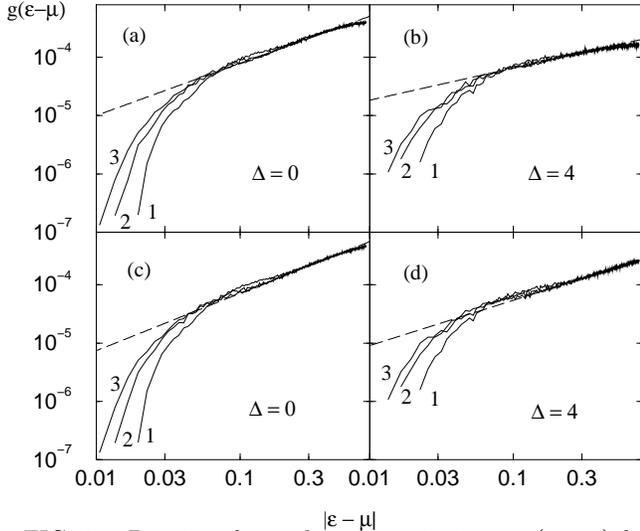}}
\caption{ 
Density of one electron excitations $g_a(\varepsilon-\mu)$
for $\varepsilon > \mu$ (a,b) and $\varepsilon < \mu$ (c,d)
obtained for the two-dimensional model (\protect{\ref{energy}}) at
$\Delta=0$ (a,c) and 4 (b,d), with $N=500$ (curves numbered 1),
1000 (2) and 1500 (3). The dashed lines are least-squares
power-law fits $g_a(\varepsilon-\mu)\sim |\varepsilon-\mu|^{\gamma}$
with $\gamma=0.9$ (a), $0.55$ (b), $0.98$ (c)
and $0.78$. Data presented in the figure are calculated as
the average over 10.000 different samples (except the case $N=1500$
and $\Delta=4$ with the average over 3700 different samples).
}
\label{fig2}
\end{figure}

According to the symmetry relation (\ref{mapp}) $g_a(\varepsilon)$ and
$g_d(\varepsilon)$ can be easily mapped to each other for any values
of $\varepsilon$ and hence all the results presented below concern the
acceptor sites only. One can expect that the width of the Coulomb gap
$\Delta \varepsilon$ and the energy scale in our model $E_0=e^2
n^{1/D}/\chi$ are of the same order of magnitude. Fig.\ref{fig1} shows
$g_a(\varepsilon-\mu)$ in the vicinity of the Fermi energy $\mu$
obtained for the two-dimensional samples with $N=1000$ and various
values of $\Delta$. As it is seen, $g_a(\varepsilon-\mu)$ depends
considerably on $\Delta$ except for a narrow window $|\varepsilon
-\mu|\lesssim 0.05$, where all data merge into some ``universal''
curve symmetric with respect to $\mu$, the curve which can be
anticipated to obey the Efros universality hypothesis (\ref{efros}). However,
a double-logarithmic plot of the ``universal'' $g_a(\varepsilon-\mu)$
(insert in the Fig.\ref{fig1}), reveals that the behavior of
$g_a(\varepsilon-\mu)$ in the ``universality'' region is not even a
power law. The width of this ``universality'' region is comparable to
the width of the region where $g_a(\varepsilon-\mu)=0$ due to the
finite size effects (for the data presented in Fig.\ref{fig1} relation
(\ref{finite1}) gives $|\varepsilon-\mu|<0.011$), so it is plausible
to suggest that the ``universal'' behavior of $g_a(\varepsilon-\mu)$
is governed by the finite-size effects. This is clearly demonstrated
in Fig.\ref{fig2} where $g_a(\varepsilon-\mu)$ are shown
for several sizes of the samples investigated. 

\begin{figure}
\epsfxsize=3.33in
\centerline{\epsfbox{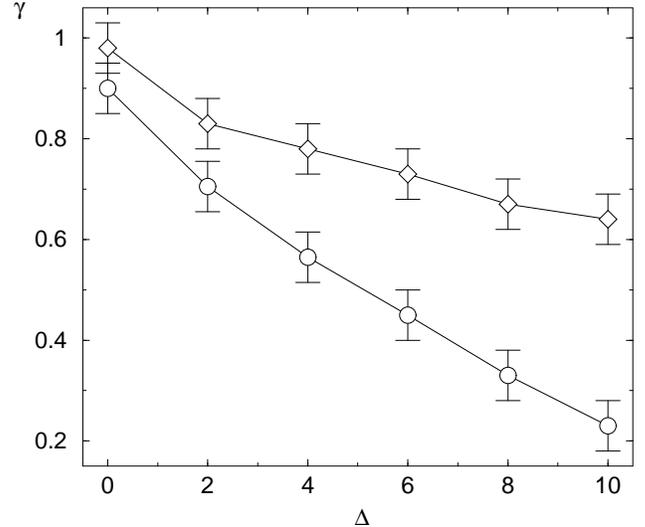}}
\caption{The exponent $\gamma$ of the power law
$g_a(\varepsilon-\mu)\sim |\varepsilon-\mu|^{\gamma}$ as a function of
the charge-transfer energy $\Delta$. The data are obtained from
least-squares fits of $g_a(\varepsilon-\mu)$ for the two-dimensional
model (\protect{\ref{energy}}) with $N=1500$ within the region $0.2
\lesssim |\varepsilon-\mu|\lesssim 0.7$. Circles represent the
positive values of $\varepsilon-\mu$ while diamonds stand for the
negative values of $\varepsilon-\mu$. Lines are guides to the eye.}
\label{fig3}
\end{figure}

The $\varepsilon$ window where finite size effects are severe, shrinks
considerably with increasing $N$ for all values of $\Delta$ we
investigated. For instance, $g_a(\varepsilon-\mu)$ for $N=500$ and
$N=1000$ at $\Delta=0$ (see Fig.\ref{fig2}a,c) merge when
$|\varepsilon -\mu| \gtrsim 0.2$ while corresponding curves for
$N=1000$ and $N=1500$ are indistinguishable already at
$|\varepsilon-\mu| \gtrsim 0.1$. The statistical noise observed for the
curves in Fig.\ref{fig2} is quite small even close to $\mu$ and
hence, the influence of insufficient large statistics on the results
obtained is excluded. Note, that the ``universal'' behavior of
$g(\varepsilon)$ in the vicinity of $\mu$ obtained for the classical
$d-a$ model (see Fig.3 in Ref.\onlinecite{dar84}) is most likely due
to the finite size effects as well.

\begin{figure}
\epsfxsize=3.33in
\centerline{\epsfbox{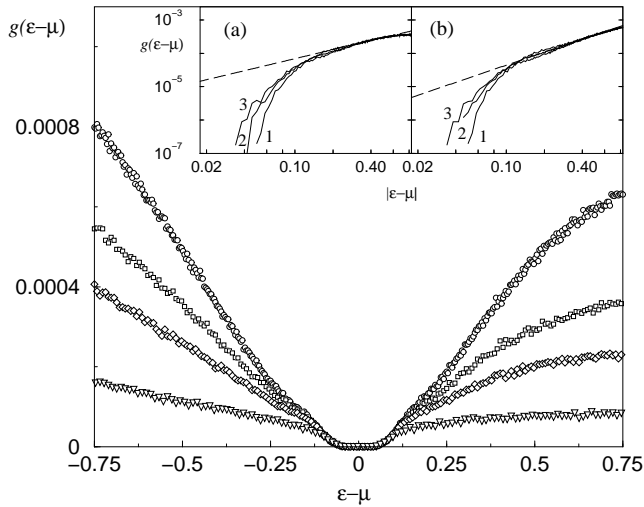}}
\caption{
Density of one electron excitations $g_a(\varepsilon-\mu)$ in
the vicinity of the Fermi energy $\mu$ obtained for
the three-dimensional model (\protect{\ref{energy}}) with $N=1000$ at
$\Delta=0$ (circles), $2$ (squares), $4$ (diamonds) and $10$
(triangles). Data points presented in the figure are calculated as
the average over 10.000 different samples. Inserts show
double-logarithmic plots of $g_a(\varepsilon-\mu)$ at $\Delta=2$,
for $N=500$ (curves numbered 1), 1000 (2) and 2000 (3), 
in the regions $\varepsilon > \mu$ (a) and for $\varepsilon < \mu$ 
(b). The dashed lines in the inserts are least-squares
power-law fits $g_a(\varepsilon-\mu)\sim |\varepsilon-\mu|^{\gamma}$
with $\gamma=1.16$ (a), $1.29$ (b),
}
\label{fig4}
\end{figure}

In the region $|\varepsilon -\mu| \gtrsim 0.2$, where the curves for
all $N$ collapse into a single curve (and where we believe the
thermodynamic limit is reached), the behavior of
$g_a(\varepsilon-\mu)$ is described by a power law
$g_a(\varepsilon-\mu)\sim |\varepsilon-\mu|^\gamma$. The deviation
from the power-law observed far away from $\mu$ ($|\varepsilon-\mu|
\gtrsim 0.7$) is due to the boundaries of the Coulomb gap which, as 
was mentioned above, are $\sim 1$ in units of $E_0$. One can see
from a comparison of the data shown in Fig.\ref{fig2} for different
$\Delta$, that the exponent $\gamma$ depends considerably on $\Delta$.
Furthermore, values of $\gamma$ in the region $\varepsilon-\mu>0$ and
those in the region $\varepsilon-\mu<0$ differ as well with this
difference increasing with increasing $\Delta$. The data for $\gamma$
obtained for the two-dimensional MCDAM are summarized in
Fig.\ref{fig3} where a significant deviation of $\gamma$ from the
value $D-1$ predicted by the hypothesis (\ref{efros}) is observed at all
values of $\Delta$ investigated except for the case $\Delta =0$ when
$\gamma \approx 1$ within the limits of statistical accuracy. Note,
that the deviation of $\gamma$ from its predicted value grows
monotonically with increasing $\Delta$. At $\Delta=10$ where the
features of the MCDAM are expected to be nearly the same as those of
the classical $d-a$ model with all the acceptors being ionized
(indeed, the degree of the acceptor ionization $C_a\sim 0.9$ for the
two-dimensional MCDAM at $\Delta=10$, see Fig.\ref{fig6} below) the
deviation from the Efros exponent is very large.

\begin{figure}
\epsfxsize=3.33in
\centerline{\epsfbox{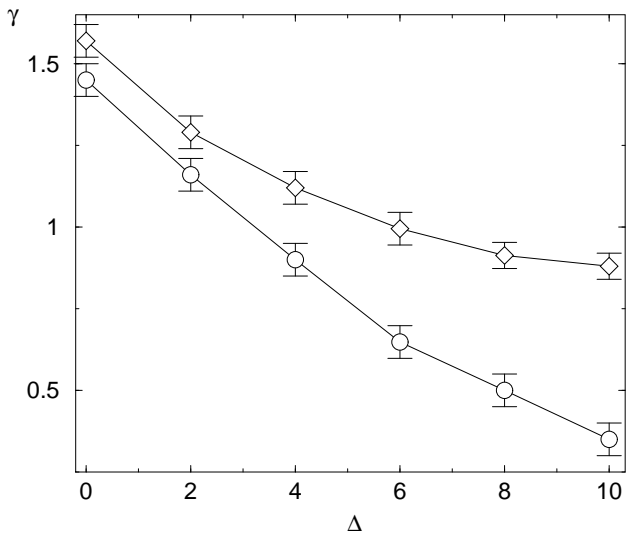}}
\caption{
The exponent $\gamma$ of the power law
$g_a(\varepsilon-\mu)\sim |\varepsilon-\mu|^{\gamma}$ 
as a function of the charge-transfer energy $\Delta$. The data 
are obtained from least-squares fits of 
$g_a(\varepsilon-\mu)$ for the three-dimensional model 
(\protect{\ref{energy}}) with $N=1000$ within the region $0.4
\lesssim |\varepsilon-\mu|\lesssim 0.8$. Circles represent the
positive values of $\varepsilon-\mu$ while diamonds stand for the
negative values of $\varepsilon-\mu$. Lines are guides to the eye.}
\label{fig5}
\end{figure}

The main results for $g_a(\varepsilon-\mu)$ obtained for the
three-dimensional MCDAM are summarized in Figs. \ref{fig4} and
\ref{fig5}. It is seen, that the behavior of $g_a(\varepsilon-\mu)$ in
three dimensions does not differ qualitatively from the behavior of
$g_a(\varepsilon-\mu)$ in two dimensions. Some quantitative
differences observed arise from the fact that at given $N$ (the
parameter which determines the amount of computer memory needed for
the calculations) the linear size of a two-dimensional sample with a
given density of sites is larger than that of a three-dimensional
sample with the same density of sites and thereby, the finite size
effects for three-dimensional samples with given $N$ are more
pronounced compared to those for the two-dimensional samples with
the same $N$. For example, the lower boundary of the region where
$g_a(\varepsilon-\mu)$ can be described by the power law
$|\varepsilon-\mu|^\gamma$ shifts towards larger $|\varepsilon-\mu|
\gtrsim 0.4$ values (see inserts in Fig.\ref{fig4}). Remarkably,
the exponent $\gamma$ does not reach the value $D-1$ predicted by the
universality hypothesis (\ref{efros}) even at $\Delta=0$ (Fig.\ref{fig5}).

\begin{figure}
\epsfxsize=3.33in
\centerline{\epsfbox{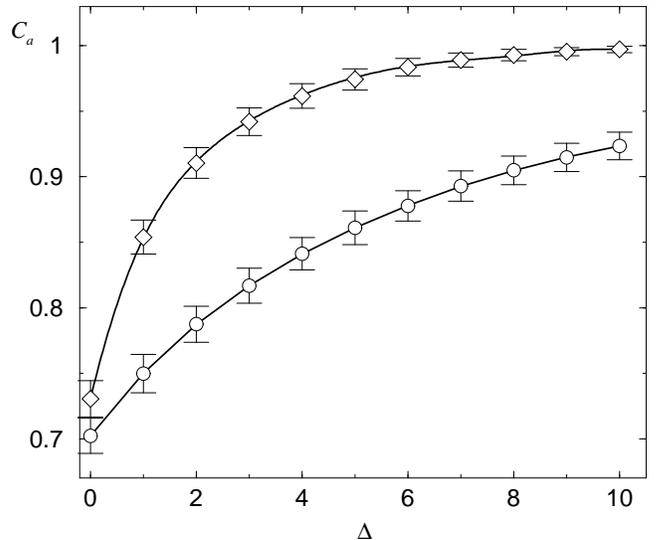}}
\caption{ The degree of acceptor ionization $C_a$ as a function of the
charge-transfer energy $\Delta$. The data are obtained for the model
(\protect{\ref{energy}}) in two (circles) and three (diamonds)
dimensions with $N=500$ as an average over 1000 different samples. The
solid lines are third-degree polynomial fits. }
\label{fig6}
\end{figure}

Unlike $g_a(\varepsilon-\mu)$ in the vicinity of the Coulomb gap, the
density of ionized acceptors $C_a$ (\ref{ioniz}) describes the
state of the entire sample and therefore reaches the thermodynamic
limit much faster than $g_a(\varepsilon-\mu)$. This allows us to
obtain quite accurate results for $C_a$ from data on a relatively
small amount of samples with $N=500$ only. Fig.\ref{fig6} shows the
variations of $C_a$ with $\Delta$ both for two and three
dimensions. In three dimensions almost all acceptors become ionized
($C_a\sim 1$) rather soon while for two
dimensions even for the largest $\Delta$ investigated around 10 \% of
the acceptors remain neutral. So, one can say, that the
three-dimensional MCDAM at $\Delta \gtrsim 7$ reduces already to the
classical $d-a$ model.  It is known that the classical $d-a$ model
exhibits in three dimensions the, so called, Coulomb fluctuational
catastrophe \cite{efros84}. For calculations on finite samples it
implies that statistical fluctuations of $\mu(\bf R)$ grow
dramatically with increasing $\Delta$ which is the case in our
calculations (see Table \ref{tab1}). Therefore, in order to reduce the
statistical noise in three dimensions, the average of
$g_a(\varepsilon-\mu)$ over a much larger (compared to $D=2$) number of
samples is needed. Note, that $\mu(\bf R)$ in both two and
three dimensions are scattered according to the Gaussian distribution
with the mean $\bar{\mu}$ obeying the relation (\ref{mudelta}).

\begin{table}
\caption{The means  $\bar{\mu}$ and standard deviations 
$\Delta \mu$ of the Fermi energy calculated for
the three-dimensional model (\protect{\ref{energy}}) with $N=1000$
and various $\Delta$.}

\vspace{1cm}

\begin{tabular}{ccccc}
%\squeezetable
&$\hspace{1cm}\Delta\hspace{1cm}$ 
&$\hspace{1cm} \bar{\mu}\hspace{1cm} $ & 
$\hspace{1cm}\Delta \mu \hspace{1cm}$&\\
\hline
&0& -0.017 & 0.100&\\
&2& -1.016 & 0.187&\\
&4& -2.0149 & 0.287&\\
&6& -3.016 & 0.392&\\
&8& -4.011 & 0.502&\\
&10& -5.018 & 0.607&\\
\end{tabular}
\label{tab1}
\end{table}

\section{DISCUSSION}

The behavior of $g_a(\varepsilon-\mu)$ calculated within the region of
the Coulomb gap for the model (\ref{energy}) is in strong
contradiction to the universality hypothesis (\ref{efros}).  Despite
the fact that $g_a(\varepsilon-\mu)$ is indeed described by the power
law $|\varepsilon-\mu|^\gamma$ in a wide range of $\varepsilon$ inside
the region of the Coulomb gap, the exponent $\gamma$ is considerably
smaller than that predicted by the hypothesis (\ref{efros}) both for
the two- and three-dimensional cases. Moreover, the exponent $\gamma$
depends significantly on $\Delta$ and is different for the cases
$\varepsilon>\mu$ and $\varepsilon<\mu$.  It is believed that
information about $g(\varepsilon)$ might be directly obtained from
tunneling and photoemission experiments \cite{white86} and recent
experiments \cite{lee99} on boron-doped silicon crystals have shown
that the density of one-electron excitations at higher energies obeys
a power-law with an exponent slightly less than $0.5$ which is in good
agreement with our results for $D=3$ and $\Delta \gtrsim 8$. However,
the non-metallic samples show around the Fermi energy a nearly quadratic
Coulomb gap, so the question arises whether our results could be
related to the intermediate asymptotic behavior observed? Here we want to
make three remarks concerning this question:

First, the power law $g_a(\varepsilon-\mu)\sim |\varepsilon-\mu|^\gamma$
is valid above a value $\varepsilon_0(N)$ below which the
finite size effects take over (Figs.\ref{fig2} and \ref{fig4}). It
seems from our results, that $\varepsilon_0(N)\rightarrow \mu$ when
$N\rightarrow \infty$. In two dimensions we were able to obtain
size-independent results down to $\varepsilon_0\sim 0.1$, i.\ e.\ for
$\sim 90 \%$ of the whole Coulomb gap, the halfwidth of which is $\sim 1$
in units of $E_0$.

Secondly, as follows from the ground-state stability relations
(\ref{stabcon}), the distance $r_{ij}$ between a neutral donor, with
an energy, say, $\varepsilon_i^1 \in [-\varepsilon,0]$ ($\varepsilon$
here is the halfwidth of a narrow band around $\mu=0$) and a charged
donor with an energy $\varepsilon_j^0 \in [0,\varepsilon]$ should be
not less that $\frac{1}{2\varepsilon}$. I.\ e., sites with energies
$\varepsilon_i^1 \in [-\varepsilon,0]$ cannot be inside a D-dimensional
sphere of radius $R_{sp}=\frac{1}{2\varepsilon}$
and with the center in a site with the energy $\varepsilon_j^0
\in [0,\varepsilon]$. Assuming that {\it all} such spheres {\it do not
intersect}, the total volume occupied by the spheres is
\begin{equation}
V_{sp} =N \times S(D) \left(\frac{1}{2\varepsilon} \right)^D \;
 \int^{\varepsilon}_0 g(\varepsilon ')d\varepsilon'
\label{assumption}
\end{equation}
where $S(D)$ is the volume of a D-dimensional sphere with the radius
equal to unity. Since $V_{sp}$ cannot exceed the total volume $V$ of
a sample ($V=N$ at $n=1$) we arrive at the inequality
\begin{equation}
 \int^{\varepsilon}_0 g(\varepsilon ')d\varepsilon '\; \leq \;
\frac{(2\varepsilon)^D}{S(D)},
\label{unequal1}
\end{equation}
which is valid for all $\varepsilon$ if
\begin{equation}
g(\varepsilon)\; \leq \; \frac{D\times 2^D}{S(D)}\; |\varepsilon|^{D-1}
\label{unequal2}
\end{equation}
The universality hypothesis (\ref{efros}) then is a limit case of
(\ref{unequal2}).  The density of sites with energies
$\varepsilon_i^1 \in [-\varepsilon,0]$ indeed decreases when
$\varepsilon \rightarrow 0$, so the assumption (\ref{assumption}) for
the spheres with {\it finite} radii seems to be plausible. However,
simultaneously $R_{sp}\rightarrow \infty$ and consequently the plausibility
of the assumption (\ref{assumption}) and thereby of the hypothesis
(\ref{efros}) becomes questionable.

And finally, the universality hypothesis (\ref{efros}) can be also
obtained as the asymptotic behavior of a non-linear integral equation
for $g(\varepsilon)$ as $\varepsilon \rightarrow 0$, the equation
which, in turn, is heuristically obtained from the stability condition
(\ref{stabcon}). The derivation of this integral equation (given, for
example, in Ref.\onlinecite{mog89}) is based on the implicit assumption
that the sites with charged donors are randomly distributed in
space according to the Poisson statistics. However, it was
unequivocally demonstrated in computer studies of the Coulomb gap
\cite{dar84} that charged donor sites with energies close to $\mu$
tend to form clusters (Ref.\onlinecite{dar84}, Fig. 6).

We conclude that $g_a(\varepsilon-\mu)$ in the region of the Coulomb
gap in model (\ref{energy}) has a power law behavior for all energies
down to $\mu$ and that the universality hypothesis of Efros
(\ref{efros}) is questionable. Note, that our results are in
contradiction not only to the universality hypothesis (\ref{efros}),
but to the inequality (\ref{unequal2}) as well. Up to now, all
exponents found are in good agreement with this inequality. E.\ g.\ in
Ref.\onlinecite{mob88} specimens of 40 000 and 125 000 sites for two-
and three-dimensional samples were investigated in the Efros' lattice
model\cite{efros76} and the power law $g_a(\varepsilon-\mu)\sim
|\varepsilon-\mu|^\gamma$ was found with $\gamma=1.2\pm 0.1$ and
$\gamma=2.6\pm 0.2$ for two and three dimensions, respectively. The
main conclusion 
\onecolumn
\begin{table}
%\squeezetable
\caption{Some donor--acceptor pairs for which the difference between the
donor and acceptor energy levels does not exceed 10 meV. $E_g$, $E_v$ 
and $E_c$ are, respectively, the energy gap, the top of the valence band
and the bottom of the conductivity band. If the solubilities of both donor
and acceptor are known, the parameter $E_0$ is calculated using the data
for the less soluble of the pair.}

\vspace{1cm}

\begin{tabular}{cccccccc}
Donor & Acceptor& \multicolumn{2}{c}{Solubility, cm$^{-3}$}&
\multicolumn{2}{c}{$E_0$, meV}& $E_j$, meV & $\Delta$, meV \\
& & min & max & min & max & &\\
\tableline
\multicolumn{8}{c}{Si ($E_g=1124$ meV, $\chi=12$)}\\
& & & & & & &  \\ 
Fe & & $1.2\times 10^{16}$ & $4\times 10^{16}$ & 0.6 &
 4  & $E_c-796 $ & 8 \\
& Zn & $2.3\times 10^{16}$ & $8\times 10^{16}$ & &
  & $E_v+320 $ & \\
& & & & & & &  \\ 
Ni & & $ 10^{18}$ & $10^{13}$ & 12 &
 $<1$  & $E_v+(160 \div 190) $ & 3.1 -- 33.1 \\
& In & $3\times 10^{17}$ & $4\times 10^{18}$ &  &
  & $E_v+156.9 $ &  \\
\tableline
\multicolumn{8}{c}{Ge ($E_g=740$ meV, $\chi=15.9$)}\\
& & & & & & &  \\ 
S & &  \multicolumn{2}{c}{no reliable data} &  &
 $ $  & $E_c-296 $ & 4 \\
& Ni & $4.8\times 10^{15}$ & $8\times 10^{15}$ & 1.5 &
 1.8 & $E_c-300 $ &  \\
\tableline
\multicolumn{8}{c}{GaAs ($E_g=1520$ meV, $\chi=12.5$)}\\
& & & & & & &  \\ 
Ti& & $2\times 10^{16}$ & -  & 3.1 & -  & $E_c-1000$ & 0 \\ 
& Fe & & & & &$E_v+520$ & 
\end{tabular}

\label{tab2}
\end{table}

\begin{multicols}{2}
\noindent of our results and those of Ref.\onlinecite{mob88} is
that Efros' lattice model\cite{efros76} can not be used to as a reliable
approximation to the classical $d-a$ model with Poisson impurity
distribution.
Energy levels of donor (acceptor) impurities are usually close to the
bottom (top) of the conduction (valence) band. Since in the most
common semiconductors the energy gap $E_g\sim 10^4$ K and $E_0\sim 20$
K, $\Delta \gg 1$ and one may ask what physical relevance does the
model (\ref{energy}) with a finite $\Delta \lesssim 10$ have, except
for being a pure academic exercise? However, in the case of deep
impurities the energy levels for some donor--acceptor pairs are
extremely close to each other not excluding even the case $\Delta =0
$\cite{enc2}. Table \ref{tab2} shows some donor--acceptor pairs with
$\Delta \lesssim 10$ in the most common semiconductors. The
solubilities of these impurities are rather low, thereby reducing the
temperature at which the Coulomb gap with features described by the
model (\ref{energy}) can be observed. Fortunately, these temperatures
are high enough ($\sim 10 \div 20$ K) for modern experimental
techniques and hence experimental observation of the Coulomb gap in
the semiconductors with deep impurities is possible to accomplish.

\section{SUMMARY}

We have studied a model of impurities in semiconductors with
infinite-range Coulomb interactions between donors, between acceptors
and between donors and acceptors. A new parameter introduced in the
model is the finite energy $\Delta$ of charge transfer between donors
and acceptors, a parameter which enables processes of ionization of
neutral impurities and of recombination of charged impurities. In the
particular case of equal amounts of donor and acceptor impurities, we
derived rigorous relations for the symmetry of the model with respect
to exchange of donor and acceptor sites. We also extended the
previously known algorithm to find the ground state including the
stability relations with respect to ionization and recombination
processes and performed computer studies of the model proposed at zero
temperature on a number of two- and three-dimensional samples with
randomly distributed $N$ donors and $N$ acceptors. We explored the
energy region around the Fermi energy $\mu$ where the Coulomb gap in
the density of one-electron excitations $g(\varepsilon)$ is
observed. The analysis of the calculated histograms $g(\varepsilon)$
revealed that the behavior of $g(\varepsilon)$ obtained from the
simulations on finite samples in the immediate neighborhood of $\mu$
is determined solely by the finite size effects. In the region where
finite size effects become negligible $g(\varepsilon)$ is described by
a power law with an exponent considerably depending on the parameter
$\Delta$ and on the sign of $\varepsilon-\mu$. Our findings challenge
the Efros universality hypothesis. Moreover, our results are in
contradiction to the main inequality (\ref{unequal2}) of which
Efros' universality hypothesis is a particular case.  We have
reexamined the heuristic derivation of the Efros hypothesis and shown
that some implicit assumptions which lead to universality are
questionable. From the analysis of experimental data on admixtures in
semiconductors we put forward possible experimental situations where
one could observe the Coulomb gap with the features being the same as
those of the model with a finite $\Delta$.

\section*{Acknowledgements}

This research was supported by The Swedish Natural Science Council
and by The Swedish Royal Academy of Sciences.

\end{multicols}

\end{document}